\documentstyle[prl,aps,multicol,psfig,epsfig]{revtex}

\begin{document}
\draft

\author{V.\ M.\ Pudalov$^{a,b}$, M.\ E.\ Gershenson$^{a}$, H.\ Kojima$^{a}$, N.\ Butch$^{a}$, E.\ M.\
Dizhur$^{c}$,\\
G.\ Brunthaler$^d$, A.\ Prinz$^d$, and G.\
Bauer$^d$}
\address{$^a$ Department of Physics and
Astronomy, Rutgers University, New Jersey 08854, USA}
\address{$^{b}$ P.\ N.\ Lebedev Physics
Institute, 119991 Moscow, Russia }
\address{$^{c}$ Institute for High Pressure Physics, Troitsk, Russia}
\address{$^d$ Institut f\"{u}r Halbleiterphysik,
Johannes Kepler Universt\"{a}t, Linz, Austria}
%\date{\today}
\title{The low-density spin susceptibility and effective mass of mobile
electrons\\
 in Si inversion layers}
\maketitle

\begin{abstract}
We studied the  Shubnikov-de Haas (SdH) oscillations in
high-mobility Si-MOS samples over a wide range of carrier
densities $n\simeq (1-50) \times 10^{11}$cm$^{-2}$, which includes
the vicinity of the apparent metal-insulator transition in two
dimensions (2D MIT). Using a novel technique of measuring the SdH
oscillations in superimposed and independently controlled parallel
and perpendicular magnetic fields, we determined the spin
susceptibility $\chi^*$, the effective mass $m^*$, and the
$g^*$-factor for mobile electrons. These quantities increase
gradually with decreasing density; near the 2D MIT, we observed
enhancement of $\chi^*$ by a factor of $\sim 4.7$.
\end{abstract}


\vspace{-0.1in}
\begin{multicols}{2}
Many two-dimensional (2D) systems exhibit an apparent
metal-insulator transition (MIT) at low temperatures as the
electron density $n$ is decreased below a critical density $n_c$
(for reviews see, e.g., Refs.~\cite{aks,amp,app}). The phenomena
of the  MIT and `metallic' conductivity in 2D attract a great deal
of interest, because it addresses a fundamental problem of the
ground state of strongly correlated electron systems. The strength
of electron-electron interactions is characterized by the ratio
$r_s$ of the Coulomb interaction energy to the Fermi energy
$\epsilon_F$. The 2D MIT is observed in Si MOSFETs at $n_c \sim
1\times 10^{11}$\,cm$^{-2}$, which corresponds to $r_s \sim 8$
(for 2D electrons in (100)-Si, $r_s = 2.63\sqrt{10^{12}{\rm
cm}^{-2}/n}$).

In the theory of electron liquid, the electron effective  mass
$m^*$, the $g^*$-factor, and the spin susceptibility $\chi^*
\propto g^*m^*$ are renormalized depending on $r_s$
\cite{abrikosov_87}. Though the quantitative theoretical results
\cite{iwamoto_91,kwon_94,chen_99} vary considerably, all of them
suggest enhancement of  $\chi^*$, $m^*$ and $g^*$ with $r_s$.
Earlier experiments \cite{fang_68,smith_72,okamoto,pan_99} have
shown growth of $m^*$ and $g^*m^*$ at relatively small $r_s$
values, pointing to a ferromagnetic type of interactions in the
explored range $1 \lesssim r_s < 6.5$. Potentially, strong
interactions might drive an electron system towards ferromagnetic
instability \cite{abrikosov_87}. Moreover, it has been suggested
that the `metallic' behavior in 2D is accompanied by a tendency to
a ferromagnetic instability \cite{finkelstein}. Thus, in relation
to the still open question of the origin of the 2D MIT, direct
measurements of these quantities in the dilute regime near the 2D
MIT are crucial.

In this Letter, we report the {\em direct} measurements of
$\chi^*$, $m^*$, and hence $g^*$ over  a wide range of carrier
densities ($1 \leq r_s \leq 8.4$), which extends for the first
time down to and across the 2D MIT. The data were obtained by a
novel technique of measuring the interference pattern of
Shubnikov-de Haas (SdH) oscillations in {\em crossed magnetic
fields}. The conventional technique of measuring $g^*m^*$ in
tilted magnetic fields \cite{fang_68,okamoto} is not applicable
when the Zeeman energy is greater than the cyclotron energy
\cite{note_okamoto}. The crossed field technique removes this
restriction and allows us to extend measurements over the wider
range of $r_s$. We find that for small $r_s$, the $g^*m^*$ values
increase slowly in agreement with the earlier data by Fang and
Stiles \cite{fang_68} and Okamoto et al. \cite{okamoto}.  For
larger values of $r_s$, $g^*m^*$ grows faster than it might be
extrapolated from the earlier data. At our highest value of
$r_s\simeq 8.4$, the measured $g^*m^*$ is greater by a factor of
4.7 than that at low $r_s$.

Our resistivity measurements were performed by ac (13Hz) technique
at the bath temperatures $0.05 - 1.6$\,K  on six (100)~Si-MOS
samples selected from 4 wafers: Si12 (peak mobility $\mu \simeq
3.4$\,m$^2$/Vs at $T=0.3$\,K), Si22 ($\mu \simeq 3.2$\,m$^2$/Vs),
Si57 ($\mu \simeq 2.4$\,m$^2$/Vs), Si6-14/5, Si6-14/10, and
Si6-14/18 ($\mu \simeq 2.4$\,m$^2$/Vs for the latter three
samples) \cite{samples}. The gate oxide thickness was $190 \pm
20$\,nm for all the samples; the 2D channel was oriented along
[011] for sample Si3-10, for all other samples - along [010]. The
crossed magnetic field system consists of two magnets, whose
fields can be varied independently. A split coil, producing the
field $B_\perp \leq 1.5T$ normal to the plane of the 2D layer, is
positioned inside the main solenoid, which creates the in-plane
field $B_{\parallel}$ up to 8T. The electron density was
determined from the period of SdH oscillations.

Typical traces of the longitudinal resistivity $\rho_{xx}$ as a
function of $B_\perp$ are shown in Fig.\,1. Due to the high
electron mobility,  oscillations were detectable down to 0.25\,T
and a large number of oscillations enabled us to extract the
fitting parameters $g^*m^*$ and $m^*$ with a high accuracy. The
oscillatory component $\delta \rho_{xx}$ was obtained by
subtracting the monotonic `background' magnetoresistance (MR) from
the $\rho_{xx}(B_{\perp})$ dependence  (the background MR is more
pronounced for lower $n$, compare Figs.\,1 and 3).

The theoretical expression for the oscillatory component of the
magnetoresistance is as follows
\cite{SdH}:
\begin{equation}
\frac{\delta\rho_{xx}}{\rho_{0}}=\sum_s A_s \cos [\pi s (
\frac{\hbar \pi n}{e B_{\perp}}- 1)] Z_s,
\end{equation}
where
\begin{equation}
A_s = 4\exp (-2 \pi^2 s \frac {k_BT_D}{\hbar \omega_c}) \frac{2
\pi^2 s k_B T/\hbar \omega_c}{\sinh (2 \pi^2 s k_B
T/\hbar \omega_c)}.
\end{equation}

Here $\rho_{0}=\rho_{xx}(B_{\perp}=0)$,
$\omega_c=eB_{\perp}/m^*m_e$ is the cyclotron frequency, $m^*$ is
the dimensionless effective mass, $m_e$ is the free electron mass,
and $T_D$ is the Dingle temperature. We take the valley degeneracy
$g_v=2$ in Eqs.~(1),\,(2) and throughout the paper. The Zeeman
term $Z_s=\cos [\pi s \hbar \pi (n_{\uparrow}-n_{\downarrow})/(e
B_{\perp})]$  reduces to a field-independent constant for
$B_{\parallel}=0$.

Application of $B_{\parallel}$ induces beating of SdH
oscillations, which are observed as a function of $B_{\perp}$. The
beat frequency is proportional to the spin polarization of the
interacting 2D electron system \cite{bychkov_82}:

\begin{equation}
P \equiv \frac{n_{\uparrow}-n_{\downarrow}}{n}=\frac{\chi^*
B_{tot}}{g_b \mu_B n}=g^*m^*\frac{eB_{tot}}{nh},
\end{equation}
where $n_{\uparrow}$ ($n_{\downarrow}$) stands for the density of
spin-up (spin-down) electrons, $g_b\simeq 2$ is the bare g-factor
for Si, and $B_{tot}=\sqrt{B_{\perp}^2+B_{\parallel}^2}$.
Equations (2),(3) imply that the spin polarization of the electron
system is linear in $B_{tot}$ (relevance of this assumption to our
experiment will be justified below).

\begin{figure}
\centerline{\psfig{figure=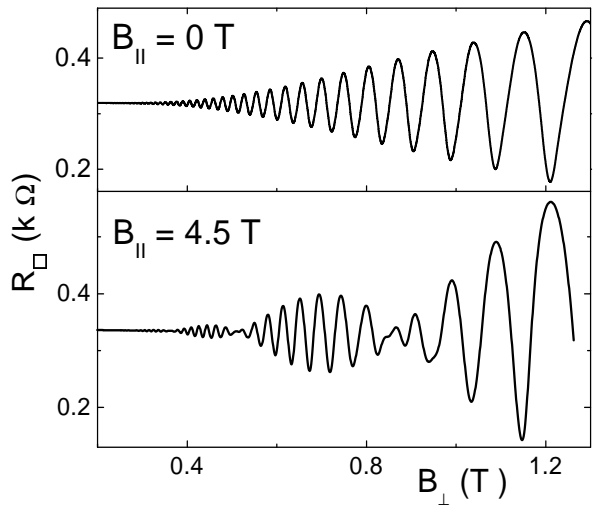,width=220pt,height=190pt,angle=0}}
\vspace{0.1in}
\begin{minipage}{3.2in} \caption{Shubnikov-de Haas
oscillations for \protect\linebreak $n=10.6\times
10^{11}$cm$^{-2}$ (sample Si6-14/10) at $T=0.35K$ and two values
of $B_{\parallel}$.} \label{Fig.1}
\end{minipage}
\end{figure}

In the experiment, we observed a well pronounced beating pattern
at a non-zero $B_{\parallel}$ (Figs.~1 and 2), in agreement with
Eq.~(1). The phase of SdH oscillations remains the same between
the adjacent beating nodes, and changes by $\pi$ through the node.
The positions of the nodes are defined solely by $g^*m^*$;
observation of several nodes enables us to determine this quantity
with a high accuracy. The positions of the nodes on the
$B_{\perp}$ axis (and, thus, the value of $g^*m^*$) are
$T$-independent at $T<1K$ within the experimental accuracy $\sim
2\%$. We have observed a non-monotonic dependence of $g^*m^*$ on
$B_{\parallel}$, which will be discussed elsewhere
\cite{Bparallel}. This dependence is more pronounced near the 2D
MIT, where $g^*m^*$ varies with $B_{\parallel}$ by $\sim 15\%$.
The data discussed below have been obtained in the fields
sufficiently low to ignore the $g^*m^*(B)$ dependence. For
example, to ensure the linear regime, at high densities ($n\sim
10^{12}$cm$^{-2}$) we have used $B_{\parallel}\simeq 0.8$\,T ,
which corresponded to the spin polarization $P\sim 1\%$; at low
densities ($n\sim 10^{11}$cm$^{-2}$), the applied
$B_{\parallel}\simeq0.1-0.2$\,T corresponded to $P\sim 5-10\%$.

\begin{figure}
\centerline{\psfig{figure=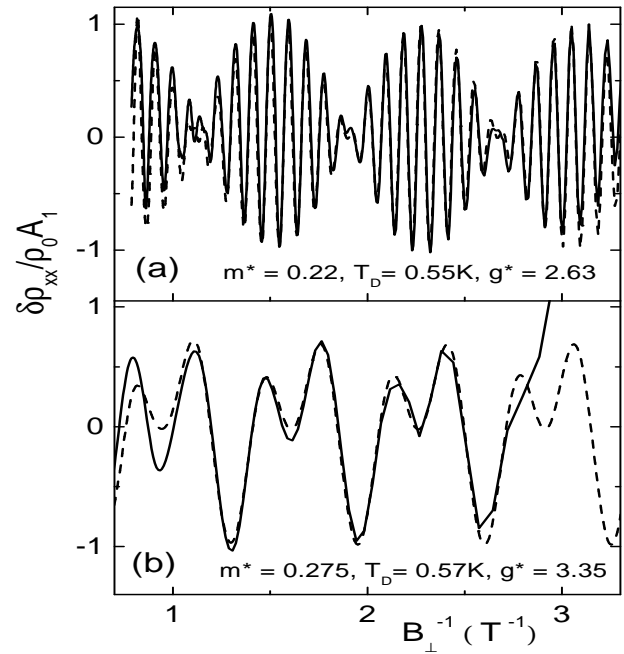,width=230pt,height=245pt}}
\vspace{0.1in}
\begin{minipage}{3.2in} \caption{Examples of  data
fitting  with Eq. (1): (a) $n=10.6 \times 10^{11}$cm$^{-2}$,
$T=0.35$\,K, $B_{\parallel}=4.5$\,T, $P=6$\%; (b) $n=2.2 \times
10^{11}$cm$^{-2}$, $T=0.4$\,K, $B_{\parallel}=3.36$\,T, $P= 34$\%.
The data for sample Si\,6-14/10 are shown as the solid lines, the
fits (with parameters shown) as the dashed lines. Both are
normalized by $A_1(B_{\perp})$. }
\end{minipage}
\label{Fig.2}
%\end{center}
\end{figure}

Comparison between the measured and calculated dependences $\delta
\rho_{xx}/ \rho_{0}$ versus $B_{\perp}$, both normalized by the
amplitude of the first harmonic $A_1$, is shown in Fig.~2 for two
carrier densities. The normalization assigns equal weights to all
oscillations. We analyzed SdH oscillations over the low-field
range $B_{\perp} \leq 1T$; this limitation arises from the
assumption in Eq.~(1) that $\hbar \omega_c \ll \epsilon_F$ and
$\delta \rho_{xx}/\rho_{0} \ll 1$. The latter condition also
allows us to neglect the inter-level interaction which is known to
enhance $g^*$ in stronger fields \cite{note_g}.

The amplitude of SdH oscillations at small $B_{\perp}$ can be
significantly enhanced by applying $B_{\parallel}$ (see Fig.~3),
which is another advantage of the cross-field technique. Indeed,
for low $B_{\perp}$ and $n$, the electron energy spectrum is
complicated by crossing of levels corresponding to different
spins/valleys. By applying $B_{\parallel}$, we can control the
energy separation between the levels, and enhance the amplitude of
low-$B$ oscillations (see Fig.~3). We have verified that
application of $B_{\parallel}$ (up to the spin polarization $\sim
20\%$) does not affect the extracted $m^*$ values (within $10\%$
accuracy), provided the sample remains in the `metallic' regime
(the insets to Fig.~3 show that the values of $m^*$ measured at
$B_{\parallel}=0$ and 3.36\,T do coincide).

\begin{figure}
\centerline{\psfig{figure=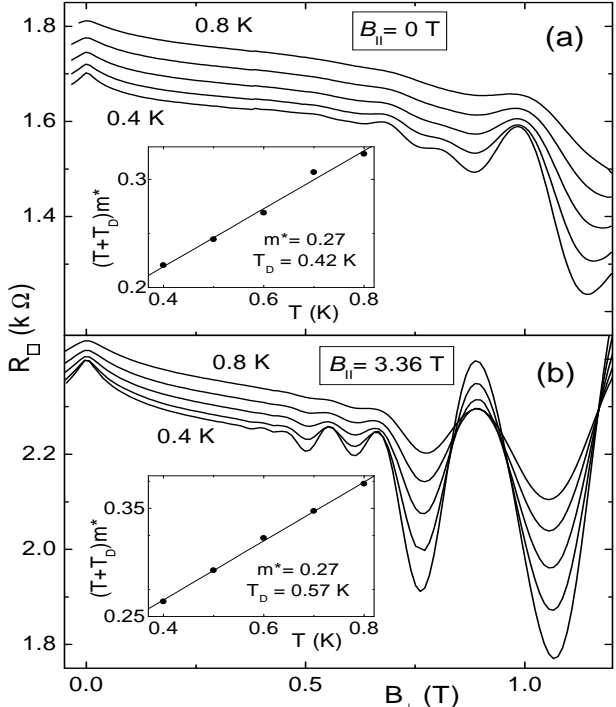,width=230pt,height=270pt}}
\begin{minipage}{3.2in}
\caption {Shubnikov-de Haas oscillations versus $B_{\perp}$ for
sample Si\,6-14/10 ($n=2.2 \times 10^{11}$cm$^{-2}$) at $T=0.4$;
$0.5$; $0.6$; $0.7$; $0.8$K: (a) $B_{\parallel}=0$ and (b)
$B_{\parallel}=3.36$\,T. The insets show the temperature
dependences of fitting parameters $(T+T_D)m^*$.} \label{Fig.3}
\end{minipage}
\end{figure}

Fitting of the  data provides us with two combinations of
parameters: $g^*m^*$ and $(T+T_D)m^*$. The measured values of
$g^*m^*$, as well as $m^*$ which are discussed below, were similar
for different samples. Figure~4\,a shows that for small $r_s$, our
$g^* m^*$ values agree with the earlier data by Fang and Stiles
\cite{fang_68} and Okamoto et al. \cite{okamoto}. For $r_s \geq
6$, $g^*m^*$ increases  with $r_s$ faster than it might be
expected from extrapolation of the earlier results \cite{okamoto}.
The MIT occurs at the critical $r_s$-value ranging from $r_c=7.9$
to 8.8 for different samples; in particular, $r_c\simeq 8.23$ for
samples Si6-14/5,10,18 which have been studied down to the MIT.

An interesting question is whether the measured dependence
$\chi^*(r_s)$ shown in Fig.\,4a represents a critical behavior as
might be expected if $\chi^*$ diverges at $n_c$. Analysis of the
$B_{\parallel}$-induced magnetoresistance led the authors of Refs.
\cite{shashkin_0007402,vitkalov_0009454} to the conclusion that
there is a ferromagnetic instability at $n = n_c$. By forcing the
critical dependence $\chi^* \propto (n/n_\chi - 1)^{-\alpha}$ to
fit our data, we found that the correlated fitting parameters
$n_\chi$ and $\alpha$ can be varied over a wide range, only weakly
affecting the standard deviation. More importantly, the period of
SdH oscillations shows that the electron states remain double spin
degenerate across the 2D MIT down to at least n =
0.98$\times$10$^{11}$\,cm$^{-2}$.  The details of this analysis,
which does not support the conclusion on spontaneous spin
polarization and divergency of $\chi^*$ at $n=n_c$, are presented
elsewhere \cite{Bparallel}.

\begin{figure}
\centerline{\psfig{figure=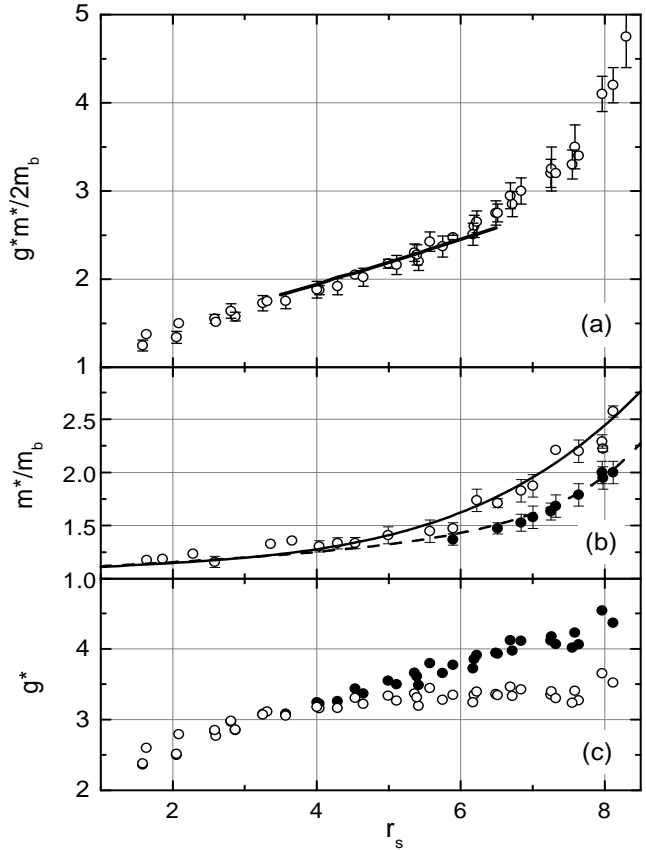,width=240pt,height=320pt,angle=0}}
\begin{minipage}{3.2in}
\caption{Parameters $g^*m^*$, $m^*$, and $g^*$ for different
samples as a function of $r_s$ (dots).
%Arrow depicts critical $r_s$-value for the sample Si6-14.
The solid line in Fig. 4a shows the data by Okamoto {\em et al.}
\protect\cite{okamoto}. The solid and open dots in Figs.\,4\,b and
4\,c correspond to two different methods of finding $m^*$ (see the
text). The solid and dashed lines in Fig.\,4b are polynomial fits
for the two dependences $m^*(r_s)$. The values of $g^*$ shown in
Fig.~4\,c were obtained by dividing the $g^*m^*$ data by the
smooth approximations of the experimental dependences $m^*(r_s)$
shown in Fig.~4\,b. } \label{Fig.4}
\end{minipage}
\end{figure}

The second combination, $(T+T_D)m^*$, controls the amplitude of
oscillations. In order to disentangle $T_D$ and $m^*$, we analyzed
the temperature dependence of oscillations over the range
$T=0.3-1.6$K (for some samples $0.4-0.8$\,K) \cite{note_temp}. The
conventional procedure of calculating the effective mass for low
$r_s$ values ($\lesssim 5$), based on the assumption that $T_D$ is
$T$-independent, is illustrated by the insets in Fig.\,3. In this
small-$r_s$ range, our results are in a good agreement with the
earlier data by Smith and Stiles \cite{smith_72}, and by Pan {\em
et al.} \cite{pan_99}. The assumption $T_D\neq f(T)$, however,
becomes dubious near the MIT, where the resistance varies
significantly over the studied temperature range; in this case,
the two parameters $T_D$ and $m^*$ become progressively more
correlated. The open dots in Fig.\,4\,b were obtained by assuming
that $T_D$ is $T$-independent over the whole explored range of
$n$: $m^*$ increases with $r_s$, and the ratio $m^*/m_b$ becomes
$\sim 2.5$ at $r_s=8$ ($m_b=0.19$ is the band mass). As another
limiting case, one can attribute the change in $R(T)$ solely to
the temperature dependence of the short-range scattering and
request $T_D$ to be proportional to $R(T)$. In the latter case,
the extracted dependence $m^*(r_s)$ is weaker (the solid dots in
Fig.\,4b). To reduce the uncertainty of $m^*$ at large $r_s$, it
is necessary to separate the effects of $T$-dependent scattering
and `smearing' of the Fermi distribution in the dependence $R(T)$;
the adequate theory is currently unavailable.

Our data shows that the combination $(T+T_D)m^*$ is almost the
same for electrons in both spin-up and spin-down subbands (e.g.,
for $n=3.76\times 10^{11}$\,cm$^{-2}$ and $B_{\parallel}=2.15$\,T
($P=20\%$), the $T_D$ values for `spin-up' and `spin-down' levels
differ by $\leq 3\%$). This is demonstrated by the observed almost
100\% modulation of SdH oscillations (see, e.g., Fig. 2\,b). Thus,
the carriers in the spin-up and spin-down subbands have nearly the
same mobility; this imposes some constraints on theoretical models
of electron transport in the 2D `metallic' state.

In conclusion, using a novel cross-field technique for measuring
SdH oscillations, we performed direct measurements of the spin
susceptibility, effective mass, and $g-$factor of conducting
electrons over a wide range of carrier densities. By studying the
temperature dependence of SdH oscillations, we disentangled three 
unknown parameters: $g^*$, $m^*$, and $T_D$, and obtained their
values in the limit of small magnetic fields. We found that both
$g^*m^*$ and $m^*$ are almost independent of the temperature over
the range $(0.2-1)$\,K. At the 2D metal-insulator transition, we
observed a finite value of the spin susceptibility, enhanced by a
factor of $\sim 4.7$ with respect to the high-density limit.

Authors are grateful to E.\ Abrahams, B. Altshuler, and D. Maslov
for discussions. The work was supported by the
NSF, ARO MURI, NATO Scientific Program, RFBR, FWF, INTAS, and
the Russian Programs \lq Physics of Nanostructures', \lq Statistical Physics', \lq
Integration', and \lq The State Support of Leading Scientific Schools'.

\end{multicols}

\begin{references}
\bibitem{note_okamoto} The
common technique of measuring $g^*m^*$ in
tilted field
 is based on detecting of disappearance of the first
harmonic of oscillations when $2g^*\mu_B B_{tot}= \hbar
eB_{\perp}/(m^*m_e)$. This technique is applicable only up to
$r_s=6.3$ \protect\cite{okamoto}.
\bibitem{samples} Samples Si6-14/5,10,18 correspond to the same
chip Si6-14 which was cooled down to 4.2K at different values of
the gate voltage $V_g=5,10,18$\,V. This resulted in different
parallel-field MR at low T (the data will be present elsewhere)
\protect\cite{Bparallel}.
\bibitem{note_temp}Below $0.3$\,K,  we observed trend of simultaneous
saturation of the temperature dependences of $\rho_0$, $\delta
\rho_{xx}$, and the dephasing rate, which is presumably caused by
electron overheating due to external electromagnetic noise.
\end{references}
\end{document}